\documentclass[conference,a4paper]{IEEEtran}
\IEEEoverridecommandlockouts

\usepackage{cite}
\usepackage{amsmath,amssymb}
\usepackage{graphicx}
\usepackage{booktabs}
\usepackage{textcomp}
\usepackage{xcolor}
\usepackage{url}
\usepackage{hyperref}
\usepackage{iftex}
\ifPDFTeX
  \usepackage[utf8]{inputenc}
  \usepackage[T1]{fontenc}
\else
  \usepackage{fontspec}
  \newif\iftelugufontok\telugufontokfalse
  \newif\iftamilfontok\tamilfontokfalse
  \newif\ifdevafontok\devafontokfalse
  \IfFontExistsTF{fonts/NotoSansTelugu-Regular.ttf}{\newfontfamily\telugufont{NotoSansTelugu-Regular.ttf}[Path=fonts/]\telugufontoktrue}{%
    \IfFontExistsTF{Noto Sans Telugu}{\newfontfamily\telugufont{Noto Sans Telugu}\telugufontoktrue}{%
      \IfFontExistsTF{Telugu Sangam MN}{\newfontfamily\telugufont{Telugu Sangam MN}\telugufontoktrue}{}}}
  \IfFontExistsTF{fonts/NotoSansTamil-Regular.ttf}{\newfontfamily\tamilfont{NotoSansTamil-Regular.ttf}[Path=fonts/]\tamilfontoktrue}{%
    \IfFontExistsTF{Noto Sans Tamil}{\newfontfamily\tamilfont{Noto Sans Tamil}\tamilfontoktrue}{%
      \IfFontExistsTF{Tamil Sangam MN}{\newfontfamily\tamilfont{Tamil Sangam MN}\tamilfontoktrue}{}}}
  \IfFontExistsTF{fonts/NotoSansDevanagari-Regular.ttf}{\newfontfamily\devanagarifont{NotoSansDevanagari-Regular.ttf}[Path=fonts/]\devafontoktrue}{%
    \IfFontExistsTF{Noto Sans Devanagari}{\newfontfamily\devanagarifont{Noto Sans Devanagari}\devafontoktrue}{%
      \IfFontExistsTF{Devanagari Sangam MN}{\newfontfamily\devanagarifont{Devanagari Sangam MN}\devafontoktrue}{}}}
  \newcommand{\textte}[1]{\iftelugufontok{\telugufont #1}\else#1\fi}
  \newcommand{\textta}[1]{\iftamilfontok{\tamilfont #1}\else#1\fi}
  \newcommand{\texthi}[1]{\ifdevafontok{\devanagarifont #1}\else#1\fi}
\fi
\usepackage{newunicodechar}
\usepackage{listings}
\usepackage{tikz}
\usetikzlibrary{positioning,shapes.geometric,arrows.meta,fit,backgrounds}
\newunicodechar{ṭ}{\d{t}}
\newunicodechar{ḍ}{\d{d}}
\newunicodechar{ṇ}{\d{n}}
\newunicodechar{ṣ}{\d{s}}
\newunicodechar{ḷ}{\d{l}}
\newunicodechar{ḻ}{\b{l}}
\newunicodechar{ɻ}{\b{r}}
\newunicodechar{ṁ}{\.{m}}
\newunicodechar{ṅ}{\.{n}}

\newcommand{\psp}{\textsc{PSP}}
\newcommand{\praxy}{\textsc{Praxy}}
\newcommand{\bups}{\textsc{BUPS}}
\newcommand{\configb}{Config B}

\begin{document}

\title{\praxy{} Voice: Voice-Prompt Recovery + \bups{} for Commercial-Class Indic TTS from a Frozen Non-Indic Base at Zero Commercial-Training-Data Cost}

\author{
  \IEEEauthorblockN{Venkata Pushpak Teja Menta}
  \IEEEauthorblockA{Praxel Ventures\\
  \texttt{pushpak@praxel.in}\\
  ORCID: \href{https://orcid.org/0009-0003-2479-9208}{0009-0003-2479-9208}}
}

\maketitle

\begin{abstract}
\sloppy
Commercial text-to-speech (TTS) systems produce near-native Indic audio today, but the best open-source bases (ResembleAI Chatterbox, Indic Parler-TTS, IndicF5) trail them by meaningful margins on measured phonological dimensions --- and the most widely adopted multilingual open-source base (Chatterbox, 23 languages) does not even tokenise Telugu or Tamil at inference. We ask: what is the minimum intervention that brings such a non-Indic-native multilingual base to commercial-class output on three tier-1 Indic languages (Telugu, Tamil, Hindi), without training a new acoustic decoder and without any commercial TTS training data? We answer with three engineering pieces combined into a single recipe: (1) \bups{}, a Brahmic Unified Phoneme Space that deterministically romanises Devanagari,\ Telugu,\ Tamil,\ Kannada,\ Bengali,\ Gujarati,\ and Malayalam to ISO-15919 so Chatterbox's Latin tokeniser can process them; (2) a LoRA adapter on \emph{only} the text-token predictor (Chatterbox's $t_3$), trained via BUPS-preprocessed Indic text with a Hindi-proxy \texttt{language\_id} on ${\sim}1{,}220$\,h of licensed Indic audio (IndicTTS, Rasa, FLEURS, Shrutilipi); (3) an inference-time voice-prompt recovery recipe that supplies a same-language reference voice clip of 8--11\,s plus three sampling overrides (exaggeration 0.7, temperature 0.6, min\_p 0.1; we call these \configb) --- recovering commercial-class acoustic output without any acoustic-decoder training. We additionally show that on Hindi, which Chatterbox natively covers, our LoRA adapter \emph{regresses} semantic accuracy, and the same \configb{} + voice-prompt recipe applied to \emph{vanilla} Chatterbox recovers commercial-class Hindi intelligibility --- giving us a two-branch deployment (LoRA branch for Te/Ta, vanilla branch for Hi) with a shared inference recipe. Evaluated on 10-utterance pilot sets with the companion \psp{}~\cite{psp2026} benchmark, our \praxy{} Voice system is on par with or slightly ahead of the best commercial baselines on three per-language headline metrics: 26.7\% retroflex collapse on Telugu (vs Sarvam Bulbul's 33.3\%; within the $n=15$-token noise band but consistently ranked first), 71\% Tamil-zha collapse (vs the commercial trio's 86\%; the single clearest per-dimension gain we observe), and 0.025 LLM-WER on Hindi (tied with Cartesia Sonic-3). All of this from a Chatterbox base that natively did not cover Telugu or Tamil at all, using zero commercial TTS training data. Finally, for intra-sentential code-mixed input we add a third branch (IndicF5 + native-script transliteration preprocessor) that drops LLM-WER on code-mix from 0.80--0.85 (raw IndicF5) to 0.14--0.27 across Hi/Te/Ta --- closing most of the gap on Te while remaining clearly behind commercial systems on Hi. We release the R6 LoRA weights under Apache-2.0, the inference code (including the code-mix preprocessor and a unified production router) under MIT, and a Gradio demo.
\end{abstract}

\begin{IEEEkeywords}
text-to-speech, Indic languages, low-resource TTS, LoRA, voice cloning, Brahmic script processing, romanisation, open-source TTS
\end{IEEEkeywords}

\section{Introduction}
\label{sec:intro}

Building a production-quality TTS system for Indic languages has historically required either (a) training a new Indic-native model from scratch on hundreds to thousands of GPU-hours (AI4Bharat Indic Parler-TTS~\cite{parler2024}, AI4Bharat IndicF5~\cite{indicf52024}, k2-fsa OmniVoice~\cite{omnivoice2026}, A2TTS~\cite{a2tts2025}), or (b) paying per-call to closed commercial systems (ElevenLabs, Cartesia Sonic-3, Sarvam Bulbul). The first path is 100--1000$\times$ beyond the reach of most product teams; the second keeps data, voices, and cost outside the team's control.

We explore a third path: \emph{minimum-intervention wrapping} of an existing frozen non-Indic-native multilingual TTS base. Our starting point is ResembleAI Chatterbox Multilingual~\cite{chatterbox2025}, an MIT-licensed open-source TTS whose full model stack (text head $t_3$, acoustic decoder $s_3\text{gen}$, voice encoder $\text{ve}$) totals 810\,M parameters, and which natively supports 23 languages --- including Hindi, but not Telugu or Tamil. Our research question is: \emph{what is the smallest change we can make to this base, at the smallest possible cost in training data and GPU-hours, that yields commercial-class output on the two tier-1 Indic languages it does not cover (Te, Ta) and maintains commercial-class output on the one it does (Hi)?}

\textbf{Our answer.} Three engineering pieces combined into one recipe:

\begin{enumerate}
  \item \bups{} (Brahmic Unified Phoneme Space): a deterministic ISO-15919~\cite{iso15919} romanisation routing layer that converts Devanagari, Telugu, Tamil, Kannada, Bengali, Gujarati, and Malayalam scripts to a Latin-script string at tokeniser-input time. Chatterbox's MTLTokenizer has no native coverage for these scripts, but it \emph{does} have rich Latin-script coverage (English, Spanish, French, Italian, Portuguese, German, Dutch, ...) --- \bups{} routes uncovered Brahmic through covered Latin at zero additional model cost.
  \item A LoRA~\cite{hu2021lora} adapter on \emph{only} the text-token predictor (Chatterbox's $t_3$ transformer), trained on BUPS-preprocessed Indic text with a Hindi-proxy \texttt{language\_id}. The acoustic decoder ($s_3\text{gen}$) and voice encoder ($\text{ve}$) remain frozen throughout. Total trainable parameters: 7.86\,M (0.97\% of the 810\,M base).
  \item A voice-prompt recovery recipe at inference: supply an 8--11\,s reference audio clip in the target language via Chatterbox's zero-shot voice-prompt interface, and apply three sampling overrides --- exaggeration 0.7, temperature 0.6, min\_p 0.1 (collectively \configb). We show empirically (§\ref{sec:configb_ablation}) that these specific overrides matter: alternatives diverge (LLM-WER $\uparrow$ 5$\times$) or under-perform.
\end{enumerate}

\textbf{Key empirical result.} On the companion \psp{} benchmark~\cite{psp2026}, this recipe puts \praxy{} Voice R6 in the commercial pack on all three tier-1 Indic languages: Telugu retroflex collapse 26.7\% (Sarvam Bulbul 33.3\%, Cartesia 50\%; \praxy{} ranks first though with overlap at small-$n$ confidence bounds, §\ref{sec:headline}); Tamil zha collapse 71\% (commercial trio 86\%; the single cleanest gain); Hindi LLM-WER 0.025 (tied with Cartesia). All without acoustic-decoder training and using zero commercial TTS training data.

\textbf{Key methodological result.} On Hindi, which Chatterbox natively covers, the LoRA adapter \emph{regresses} semantic accuracy (LLM-WER 0.025 → 0.334 when the LoRA is engaged). Interpreting this as a scope-of-method signal, we deploy a two-branch architecture for pure-script inputs: LoRA branch for Te/Ta (where the base lacks native coverage and \bups{} routing matters); vanilla Chatterbox branch for Hi (where the base already works). Both pure-script branches share the same voice-prompt + \configb{} inference recipe; code-mixed inputs route to a third branch (\S\ref{sec:codemix_routing}). The negative control on Hindi is part of the contribution: it demonstrates that \bups{} + LoRA's scope is precisely \emph{languages the base does not natively cover}, not a universal solution.

\textbf{Contributions.}
\begin{enumerate}
  \item \bups{}, a deterministic ISO-15919 routing scheme for uncovered Brahmic scripts in Latin-tokeniser TTS bases (§\ref{sec:bups}). Reference implementation under MIT.
  \item A minimum-intervention LoRA adaptation strategy that targets only the text head of a frozen multilingual TTS, combined with Hindi-proxy \texttt{language\_id} conditioning to ride the base's closest supported acoustic manifold (§\ref{sec:lora}).
  \item A voice-prompt recovery recipe (BYOR + \configb) that achieves commercial-class acoustic output at inference without acoustic-decoder training, with an empirical ablation (§\ref{sec:configb_ablation}) across three parameter configurations.
  \item A two-branch pure-script routing architecture (§\ref{sec:routing}) with a negative control on Hindi that demarcates the LoRA-adaptation method's scope.
  \item A code-mix branch (§\ref{sec:codemix_routing}, §\ref{sec:codemix_results}) that pairs IndicF5 with a native-script transliteration preprocessor; LLM-WER drops from 0.80--0.85 (raw IndicF5) to 0.14--0.27 across Hi/Te/Ta. Independent of the LoRA contribution.
  \item An Apache-2.0 release of the R6 LoRA weights at \texttt{huggingface.co/Praxel/\allowbreak praxy-voice-r6} plus MIT inference code at \texttt{github.com/praxelhq/\allowbreak praxy} (including the unified production router and the code-mix preprocessor) and a Gradio demo on HF Spaces.
\end{enumerate}

\section{Related Work}
\label{sec:related}

\textbf{Open-source multilingual TTS.} Chatterbox Multilingual~\cite{chatterbox2025} is our base: MIT-licensed, 23 languages, zero-shot voice cloning. OmniVoice~\cite{omnivoice2026}, released March 2026, offers 600+ languages via a from-scratch 581k-hour diffusion-language-model architecture; our approach is complementary (minimum wrapping vs from-scratch retraining). VoxCPM2~\cite{voxcpm2} is a 2B tokeniser-free diffusion TTS. Indic Parler-TTS~\cite{parler2024} and IndicF5~\cite{indicf52024} are purpose-built Indic TTS from AI4Bharat; we include Indic Parler-TTS as a baseline (§\ref{sec:baselines}). A2TTS~\cite{a2tts2025} is a 2025 diffusion-based Indic TTS for low-resource speaker adaptation covering Bengali, Gujarati, Hindi, Marathi, Malayalam, Punjabi, Tamil --- notably not Telugu. Our approach differs in both the input (we adapt an existing non-Indic base) and the training budget (we adapt in LoRA, they train from scratch).

\textbf{LoRA for TTS.} Parameter-efficient fine-tuning via LoRA~\cite{hu2021lora} has become standard for TTS personalisation, with prior applications focusing on speaker cloning, emotion control, and style arithmetic. To our knowledge, LoRA has not been specifically applied as a \emph{language-extension} mechanism for a multilingual base that lacks the target language entirely.

\textbf{Brahmic script romanisation.} The ISO-15919~\cite{iso15919} standard defines a lossless romanisation of Indic scripts; the \texttt{indic-transliteration}~\cite{indictransliteration} package implements the mapping. \bups{} is the first application we know of that uses romanisation as a \emph{routing mechanism} for TTS, turning a Latin-tokenised non-Indic base into an Indic-capable one.

\textbf{Voice cloning at inference.} Chatterbox, like several contemporary systems~\cite{f5tts2024,voxcpm2}, exposes a zero-shot voice-prompt interface that conditions the acoustic decoder on 3--20\,s of reference audio. Our contribution here is not the voice-prompt mechanism itself but the specific \emph{recovery recipe} (\configb{} sampling + same-language reference) that upgrades an otherwise adequate Te/Ta acoustic output to commercial-class.

\textbf{Accent evaluation for Indic TTS.} Our evaluation uses \psp{}~\cite{psp2026} (companion paper), a six-dimensional phonological accent benchmark for Indic TTS. PSR~\cite{psr2026} is a contemporary rule-based phonological benchmark for English accents.

\section{Method}
\label{sec:method}

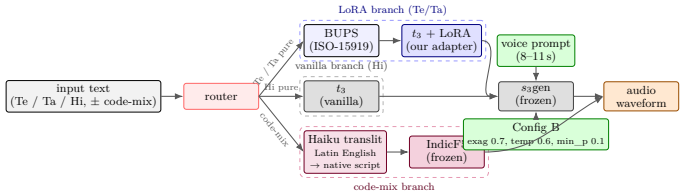
\begin{figure}[t]
\centering
\resizebox{\linewidth}{!}{%
\begin{tikzpicture}[
  font=\small,
  node distance=4mm and 6mm,
  box/.style={draw, rounded corners=2pt, align=center, minimum height=7mm,
              inner sep=3pt, minimum width=20mm},
  input/.style={box, fill=gray!10},
  trained/.style={box, fill=blue!12, draw=blue!50!black},
  frozen/.style={box, fill=gray!25, draw=gray!50!black},
  io/.style={box, fill=orange!18, draw=orange!60!black},
  shared/.style={box, fill=green!15, draw=green!50!black},
  cmstyle/.style={box, fill=purple!12, draw=purple!60!black},
  arrow/.style={-Latex, thick, gray!70!black},
]
\node[input] (text) {input text\\\small(Te / Ta / Hi, $\pm$ code-mix)};
\node[box, right=6mm of text, draw=red!60, fill=red!8] (router) {router};

\node[box, right=12mm of router, yshift=15mm, fill=blue!6] (bups) {\bups{}\\(ISO-15919)};
\node[trained, right=of bups] (t3lora) {$t_3$ + LoRA\\\small(our adapter)};

\node[frozen, right=12mm of router] (t3van) {$t_3$\\\small(vanilla)};

\node[cmstyle, right=12mm of router, yshift=-15mm] (translit) {Haiku translit\\\scriptsize Latin English\\[-1pt]\scriptsize $\to$ native script};
\node[frozen, right=of translit, fill=purple!18, draw=purple!60!black] (indicf5) {IndicF5\\\small(frozen)};

\node[frozen, right=64mm of router] (s3gen) {$s_3\text{gen}$\\\small(frozen)};
\node[shared, above=4mm of s3gen] (voice) {voice prompt\\\small(8--11\,s)};
\node[shared, below=4mm of s3gen, yshift=2mm] (configb) {\configb\\\scriptsize exag 0.7, temp 0.6, min\_p 0.1};

\node[io, right=8mm of s3gen] (wav) {audio\\waveform};

\draw[arrow] (text) -- (router);
\draw[arrow] (router.east) -- node[above, sloped, font=\scriptsize]{Te / Ta pure} (bups.west);
\draw[arrow] (bups) -- (t3lora);
\draw[arrow] (t3lora.east) to[out=0, in=180] (s3gen.west);
\draw[arrow] (router.east) -- node[above, font=\scriptsize]{Hi pure} (t3van.west);
\draw[arrow] (t3van.east) to[out=0, in=180] (s3gen.west);
\draw[arrow] (router.east) -- node[below, sloped, font=\scriptsize]{code-mix} (translit.west);
\draw[arrow] (translit) -- (indicf5);
\draw[arrow] (indicf5.east) to[out=0, in=210] (wav.west);
\draw[arrow] (voice) -- (s3gen);
\draw[arrow] (configb) -- (s3gen);
\draw[arrow] (s3gen) -- (wav);

\begin{scope}[on background layer]
\node[fit=(bups)(t3lora), draw=blue!40, dashed, rounded corners, inner sep=3pt,
      label={[blue!50!black, font=\footnotesize]above:LoRA branch (Te/Ta)}] {};
\node[fit=(t3van), draw=gray!60, dashed, rounded corners, inner sep=3pt,
      label={[gray!50!black, font=\footnotesize]above:vanilla branch (Hi)}] {};
\node[fit=(translit)(indicf5), draw=purple!40, dashed, rounded corners, inner sep=3pt,
      label={[purple!50!black, font=\footnotesize]below:code-mix branch}] {};
\end{scope}
\end{tikzpicture}}
\caption{\praxy{} Voice three-branch inference pipeline. Te/Ta pure-script routes through \textbf{LoRA branch} (\bups{} romaniser $\to$ LoRA-adapted Chatterbox $t_3$); Hi pure-script routes through \textbf{vanilla branch} (unchanged Chatterbox $t_3$); both converge on frozen $s_3\text{gen}$ with the voice-prompt + \configb{} recipe. Code-mixed inputs (any target language with $\geq$1 Latin word $\geq$2 chars) route through the \textbf{code-mix branch} (\S\ref{sec:codemix_routing}): a Haiku-driven native-script transliteration preprocessor feeds AI4Bharat IndicF5~\cite{indicf52024} (frozen, char-level) zero-shot. Only the blue $t_3$ + LoRA box has trained weights in this work; everything else is frozen base or zero-shot inference.}
\label{fig:pipeline}
\end{figure}

Figure~\ref{fig:pipeline} summarises the three-branch inference pipeline. Training applies only to the LoRA branch (Te/Ta pure-script); the vanilla branch (Hi pure-script) is unchanged Chatterbox; the code-mix branch is zero-shot IndicF5 with a transliteration preprocessor (\S\ref{sec:codemix_routing}). The LoRA and vanilla branches share the inference-time voice-prompt + \configb{} recipe.

\subsection{\bups{}: Brahmic Unified Phoneme Space}
\label{sec:bups}

\textbf{Motivation.} Chatterbox's MTLTokenizer covers 23 languages (ar, da, de, el, en, es, fi, fr, he, hi, it, ja, ko, ms, nl, no, pl, pt, ru, sv, sw, tr, zh). Passing raw Telugu (e.g., ``\textte{నేను ఇవాళ బాగున్నాను}'') or Tamil Devanagari yields either ValueError (\texttt{Unsupported language\_id}) for \texttt{te}/\texttt{ta}, or byte-BPE junk tokens for those scripts under any other \texttt{language\_id}. The model cannot learn useful token associations for uncovered scripts.

\textbf{The insight.} Chatterbox has dense Latin-script coverage across eight languages (en, es, fr, it, pt, de, nl, tr). If we can translate uncovered Brahmic scripts to Latin in a way that preserves phonological information, the tokeniser can process them via its existing Latin paths. ISO-15919~\cite{iso15919} is precisely such a romanisation --- a lossless, deterministic mapping from every Brahmic script codepoint to a Latin-with-diacritics equivalent.

\textbf{Definition.} Given input text $s$, \bups{} performs:
\begin{enumerate}
  \item Script-run segmentation of $s$ into maximal single-script spans, using Unicode block ranges: Devanagari U+0900--097F, Bengali U+0980--09FF, Gujarati U+0A80--0AFF, Tamil U+0B80--0BFF, Telugu U+0C00--0C7F, Kannada U+0C80--0CFF, Malayalam U+0D00--0D7F.
  \item ISO-15919 transliteration of each Brahmic span to Latin (via \texttt{indic-transliteration}~\cite{indictransliteration}). Non-Brahmic spans (Latin, digits, punctuation) pass through unchanged.
  \item Concatenation of the transliterated runs back into a single Latin-dominant string.
\end{enumerate}

\textbf{Example.} ``\textte{మా} CEO \textte{ఈ} quarter \textte{కి మంచి} presentation \textte{ఇచ్చారు}'' (code-mixed Telugu--English) $\to$ ``mā CEO ī quarter ki maṁci presentation icchāru''. English spans preserved; Telugu spans romanised; phonemes preserved via diacritic marks. Chatterbox's Latin tokeniser handles the result cleanly.

\subsection{LoRA Adaptation of the Text Head}
\label{sec:lora}

\textbf{Target.} Only Chatterbox's $t_3$ transformer (the text-token predictor), and within it only the attention projections (\texttt{q\_proj}, \texttt{k\_proj}, \texttt{v\_proj}, \texttt{o\_proj}), wrapped via HuggingFace's PEFT library~\cite{peft}. Our training driver builds on Ahmed-Ezzat's open LoRA framework for Chatterbox~\cite{chatterbox_ft_ahmed} for the dataset and loss scaffolding. Rank 32, alpha 64, dropout 0.05, no bias. Yields 7.86\,M trainable parameters out of the 810\,M-parameter base (0.97\%). The acoustic generator ($s_3\text{gen}$) and voice encoder ($\text{ve}$) stay frozen.

\textbf{Input format.} BUPS-preprocessed Indic transcripts (§\ref{sec:bups}) with a Hindi-proxy \texttt{language\_id}. Chatterbox's 23-language roster covers Hindi but not other Brahmic languages; by conditioning Te/Ta on \texttt{hi} we route them through the closest-matched acoustic manifold the base already knows. Preliminary experiments (Round 1, 2) with \texttt{language\_id=te} directly yielded \texttt{ValueError} and unstable training; the Hindi-proxy path was the necessary fix.

\textbf{Training data.} Approximately $1{,}220$\,h of licensed Indic speech (broken down below; figures are upper-bounded by post-UTMOS filtering, which drops ${\sim}5\%$ of clips):
\begin{itemize}
  \item IndicTTS~\cite{indictts2022}: $\sim$15 h Te, $\sim$26 h Ta, $\sim$15 h Hi.
  \item Rasa~\cite{ai4bharat2023rasa}: $\sim$20 h Hi emotion-labelled.
  \item FLEURS~\cite{conneau2022fleurs}: $\sim$5 h per language.
  \item Shrutilipi (Te/Ta/Hi subsets, filtered to clips $<$ 20\,s): $\sim$150 h Te, $\sim$700 h Hi, $\sim$280 h Ta.
\end{itemize}
All CC-BY-4.0 compatible. No commercial TTS training data is used.

\textbf{Hyperparameters.} bf16 mixed precision; AdamW $\beta=(0.9,0.95)$, weight decay 0.01; cosine LR schedule with 500-step linear warmup and peak LR $3\times10^{-6}$; batch size 16, grad-accum 1, grad-clip 0.5; total 8\,000 steps on a single A100-80GB ($\sim$11\,h wall-clock, $\sim$\$45 compute). Early training rounds at peak LR $2\times10^{-5}$ diverged on a single-language Telugu mix around step 3\,600 --- the quarter-LR setting combined with a divergence-abort heuristic (abort if EMA-loss rises $>$5\% for two consecutive save points) was the stable regime we converged on. 
\subsection{Voice-Prompt Recovery and \configb}
\label{sec:configb}

\textbf{Problem.} Even after \bups{} + LoRA, straightforward inference on Te with default Chatterbox sampling parameters (exaggeration 0.5, temperature 0.8, min\_p 0.05, no reference audio) yields intelligible-but-flat output: correct words, non-native cadence, ``foreigner accent'' per native listener feedback. The LoRA-adapted text head produces reasonable speech tokens, but the frozen acoustic decoder has no native-Indic prior to draw on.

\textbf{The recipe.}
\begin{enumerate}
  \item Supply an 8--11\,s reference audio clip in the target language via Chatterbox's \texttt{audio\_prompt\_path} interface. The voice-encoder extracts a speaker-and-prosody embedding that conditions the decoder on same-language acoustic characteristics.
  \item Override three decoder sampling parameters: \textbf{exaggeration = 0.7} (up from 0.5; increases prosodic contour), \textbf{temperature = 0.6} (down from 0.8; tightens distribution), \textbf{min\_p = 0.1} (up from 0.05; filters low-probability tokens that cause phoneme drift). We call this triple \configb.
\end{enumerate}

\textbf{What this recipe does.} The voice prompt supplies native-Indic acoustic context to a frozen-on-English-prior decoder; \configb{} sampling keeps the decoder on-distribution for that context. The LoRA'd text head carries the adaptation cheaply. Everything expensive (acoustic decoder weights) stays frozen.

\textbf{Acoustic-decoder adaptation as deferred work.} Ideally one would LoRA-adapt $s_3\text{gen}$ too; on A100-80GB its flow-matching forward+backward does not fit at any meaningful batch size with our loss definition (§\ref{sec:failure_mode} details six attempts with progressively smaller batches). H100-class hardware or gradient checkpointing would unblock it; voice-prompt recovery is the inference-time substitute we ship today.

\textbf{How \configb{} was chosen.} The three sampling overrides came from a compact three-configuration sweep (Config A / B / C, §\ref{sec:configb_ablation}) on the Telugu pilot, with final selection informed by a native-Telugu listener's ear test on category-stratified utterances (declarative / interrogative / emotional / long-narrative). We did not do a dense grid search; this is a practical engineering finding, not a hyperparameter-optimisation result, and we report it as such.

\subsection{Language-Specific Routing}
\label{sec:routing}

We deploy a two-branch architecture for pure-script inputs, augmented by a code-mix-specific third branch (§\ref{sec:codemix_routing}):

\begin{center}
\resizebox{\columnwidth}{!}{%
\begin{tabular}{@{}ll@{}}
\toprule
\textbf{Input class} & \textbf{Branch} \\
\midrule
Te (pure script) & LoRA + BUPS + Hi-proxy \texttt{lang\_id} \\
Ta (pure script) & LoRA + BUPS + Hi-proxy \texttt{lang\_id} \\
Hi (pure script) & vanilla Chatterbox (no LoRA, no BUPS) \\
Te / Ta / Hi (code-mix) & native-script transliteration $\to$ IndicF5 (\S\ref{sec:codemix_routing}) \\
\bottomrule
\end{tabular}}
\end{center}

The pure-script branches share the voice-prompt + \configb{} inference recipe (§\ref{sec:configb}). The pure-script routing rule is one line of deployment code: \texttt{if lang in \{te, ta\}: route=lora else: route=vanilla}; code-mix detection adds one further branch described in §\ref{sec:codemix_routing}.

We show empirically (§\ref{sec:scope_control}) that the pure-script routing is not arbitrary: the LoRA branch \emph{regresses} Hindi semantic accuracy, and the vanilla branch recovers it to commercial-class. The negative control is integral to the contribution.

\subsection{Code-Mix Routing via Native-Script Transliteration}
\label{sec:codemix_routing}

The pure-script branches handle one input class poorly: intra-sentential code-mix (e.g., the Hindi sentence ``\textsf{\d{m}aiṁne WhatsApp pe message kiyā but notification nahīṁ āyā}''), which dominates Indian online text. Both branches degrade on it. The LoRA branch fails because BUPS romanises English spans into Indic-phonetic readings (English ``CEO'' becomes ``\textsf{kīo}''); the vanilla branch fails because Chatterbox's per-utterance \texttt{language\_id} forces English chunks through a Hindi-conditioned acoustic manifold.

A third branch treats code-mix as a \emph{tokeniser-input distribution} problem rather than a model problem.

\textbf{Backbone substitution.} Code-mix inputs route to AI4Bharat IndicF5~\cite{indicf52024}, a flow-matching DiT TTS with a character-level tokeniser and no \texttt{language\_id} input. The architecture does not impose Chatterbox's single-language-per-utterance constraint. IndicF5 was pre-trained on 1{,}417\,h of multilingual Indic speech (IndicVoices-R, Rasa, IndicTTS, LIMMITS), all Indic-only audio. Our use is zero-shot, no further training.

\textbf{Native-script transliteration of English spans.} IndicF5's character-level tokeniser admits Latin codepoints, but the pre-training corpus contains no Latin-script audio. At inference, raw Latin spans inside Indic input are silently dropped: the model has no acoustic mapping for them. Native Indian publishers (Bollywood subtitles, Indian news tickers, and the Sarvam Bulbul training data per their February 2026 technical disclosure~\cite{sarvam_bulbul2026}) avoid this by transliterating English brand and tech words to native-script phonetic spelling: ``WhatsApp'' as ``\texthi{व्हाट्सऐप}'', ``message'' as ``\texthi{मैसेज}'', ``CEO'' as ``\texthi{सीईओ}''. We perform this preprocessing via a small instruction-tuned LM (Anthropic Claude Haiku 4.5) with a tight system prompt that (a) preserves all native-script characters and word order verbatim, (b) replaces every Latin alphabetic word with its native-script phonetic spelling, (c) keeps numbers and punctuation unchanged. We use \texttt{temperature=0} and cache by SHA-256 of the input string, so the call is deterministic and re-runs are free; the full prompt and cache file are in the public release repository. A per-utterance first call costs roughly \textasciitilde\$0.02.

The detection rule that triggers this branch is one regex: any utterance containing at least one Latin alphabetic word of length $\geq 2$. Single-letter acronym fragments and digits are handled by the unified Indic number normaliser instead. The combined recipe (transliteration $\to$ IndicF5) adds a single inference path to the deployment matrix and is independent of the LoRA contribution above.

\section{Experimental Setup}
\label{sec:setup}

\subsection{Evaluation Benchmark}
\label{sec:psp_summary}
We evaluate with \psp{} (Phoneme Substitution Profile)~\cite{psp2026}, the companion paper's six-dimensional phonological accent benchmark for Indic TTS. Since the reader should be able to interpret our results without fetching the companion paper, we summarise \psp{} here.

Four of the six dimensions are \emph{per-phoneme} probes. For each probe, \psp{} (a) force-aligns the synthesised audio against a Wav2Vec2-XLS-R~\cite{babu2021xlsr} layer-9 char-CTC aligner, (b) extracts the XLS-R embedding at each target-phoneme span, and (c) measures the cosine similarity of that embedding to two native-speaker prototype centroids --- the \emph{native} centroid and the \emph{non-native substitute} centroid --- derived from 500 Indic-native clips per language. A token is said to have \emph{collapsed} if its similarity to the substitute centroid exceeds its similarity to the native centroid. The four dimensions are: retroflex collapse rate (RR: native retroflex \d{t}, \d{d}, \d{n}, \d{s}, \d{l} collapsing to dental cognates), aspiration fidelity (AF: aspirated stops retaining breath), length fidelity (LF: long/short vowel ratios matching the $\sim$1.9$\times$ native prior), and Tamil-zha fidelity (ZF: the retroflex approximant /\b{r}/, Tamil letter \textta{ழ}).

The remaining two dimensions are \emph{corpus-level distributional} distances: FAD (Fr\'echet Audio Distance on XLS-R utterance-level embeddings, against a 1\,000-clip native reference distribution per language) and PSD (Prosodic Signature Divergence: Fr\'echet distance on a 5-D prosody vector --- pitch range, log-$F_0$ mean, speech rate, nPVI, log-duration --- against a 500-clip native reference).

Per-phoneme probes have a non-trivial native-audio noise floor ($\sim$46\% retroflex collapse on native Telugu audio, per the calibration study in~\cite{psp2026}), so we interpret per-phoneme numbers as \emph{relative rankings across systems} and reserve absolute interpretation for FAD and PSD.

\subsection{Intelligibility Metrics}
We additionally report: literal WER, LLM-WER (Qwen-2.5-72B semantic judge via OpenRouter), LLM-CER, and intent-preservation rate. All use the same STT stack (\texttt{vasista22/whisper-\{te,hi,ta\}-large-v2}, IndicWhisper-family) across every system, ensuring apples-to-apples comparison independent of STT bias.

\subsection{Test Sets}
\label{sec:test_sets}
\psp{} v1 10-utterance pilot sets per language, stratified by category (declarative, interrogative, emotional, long-agglutinative, numbers/entities, formal, colloquial, phonetically-tricky, long-narrative). For the code-mix branch evaluation (§\ref{sec:codemix_results}) we use a parallel 10-utterance code-mix smoke set per language at 25--35\% English-token density, stratified by topical category (tech, office, food, travel, money). Each utterance is synthesised with a single voice for \praxy{} ($n=10$ wavs per lang) and two voice genders for commercial systems ($n=20$ wavs per lang).

\subsection{Baselines}
\label{sec:baselines}
\begin{itemize}
  \item \textbf{Commercial}: ElevenLabs v3, Cartesia Sonic-3, Sarvam Bulbul ``bulbul:v3''. All via their public APIs; trial-tier accounts, no special credits.
  \item \textbf{Open-source}: Indic Parler-TTS~\cite{parler2024} (zero-shot), Chatterbox vanilla~\cite{chatterbox2025} (no LoRA, no BUPS) as our own reference and as the Hi-branch deployment model, AI4Bharat IndicF5~\cite{indicf52024} (zero-shot) as a second open-source point on the same smoke sets and as the back-end of our code-mix branch (§\ref{sec:codemix_routing}).
\end{itemize}

\subsection{References for the Voice-Prompt Branch}
For reproducibility, our \praxy{} inference runs use voice prompts drawn from the commercial systems' own smoke-set outputs (single longest clean clip per system, 8--11\,s): Sarvam-Te-female-9s, Sarvam-Ta-male-11s, Cartesia-Hi-female-6s. This is a v1 evaluation choice --- these clips are re-synthesised text that the commercial systems produce in a reproducible way. For production deployment (and the accompanying Gradio demo), the BYOR pattern lets the user supply any 8--20\,s same-language clip; we verified on an English-speaking author's memo (49\,s) that cross-language prompts hurt acoustic distribution, motivating the same-language constraint.

\section{Results}
\label{sec:results}


\subsection{Headline Results}
\label{sec:headline}
Table~\ref{tab:headline} shows the full \psp{} + intelligibility picture across all three languages. On Telugu, \praxy{} retroflex collapse is 26.7\% vs Sarvam Bulbul's 33.3\% --- a 1-token absolute difference on $n=15$ retroflex tokens, within small-$n$ noise, though consistently the lowest-ranked number. On Tamil, \praxy{} Tamil-zha collapse is 71\% vs the commercial trio's 86\% --- a 1-token difference on $n=7$ zha tokens (the same small-$n$ caveat applies, though the direction is the clearest per-dimension gain we observe). On Hindi, \praxy{} LLM-WER (0.025) ties Cartesia Sonic-3.

\begin{table*}[t]
\centering
\caption{Three-language headline \psp{} + intelligibility results. \praxy{} systems use our voice-prompt + \configb{} recipe; Te/Ta via the LoRA branch, Hi via the vanilla branch. Best value per column per language in bold; where \praxy{} ties the best commercial, both are bold. $n=10$ for each \praxy{} row, $n=20$ for commercial. ZF (Tamil-zha) applies only to Tamil. Dashes indicate unmeasured cells (trial-tier rate-limits capped LLM-WER scoring on some commercial rows; they are not unsatisfactory results).}
\label{tab:headline}
\small
\resizebox{\columnwidth}{!}{%
\begin{tabular}{llcccccc}
\toprule
\textbf{Lang} & \textbf{System} & FAD $\downarrow$ & PSD $\downarrow$ & RR $\downarrow$ & ZF $\downarrow$ & LLM-WER $\downarrow$ & Intent $\uparrow$ \\
\midrule
Te & Sarvam Bulbul & \textbf{250.4} & \textbf{11.1} & 33.3\% & --- & \textbf{0.029} & 0.90 \\
Te & \textbf{\praxy{} R6 + Sarvam-Te-ref} & 291.3 & 13.1 & \textbf{26.7\%} & --- & 0.033 & 0.90 \\
Te & ElevenLabs v3 & 328.9 & 154.4 & 40.0\% & --- & 0.041 & 0.85 \\
Te & Cartesia Sonic-3 & 458.1 & 33.8 & 50.0\% & --- & 0.029 & 0.90 \\
Te & Indic Parler-TTS & 325.0 & \textbf{10.4} & 33.3\% & --- & 0.144 & 0.74 \\
Te & IndicF5 (zero-shot) & --- & --- & --- & --- & 0.079 & 0.90 \\
\midrule
Ta & Sarvam Bulbul & \textbf{200.3} & 72.3 & 70.5\% & 85.7\% & --- & --- \\
Ta & ElevenLabs v3 & 239.4 & 253.7 & 69.2\% & 85.7\% & --- & --- \\
Ta & Cartesia Sonic-3 & 404.3 & 181.0 & 69.2\% & 85.7\% & --- & --- \\
Ta & Indic Parler-TTS & 233.1 & \textbf{27.1} & \textbf{64.3\%} & \textbf{61.5\%} & --- & --- \\
Ta & \praxy{} R6 + Sarvam-Ta-ref & 276.0 & 71.2 & 69.2\% & 71.4\% & 0.041 & 0.90 \\
\midrule
Hi & \textbf{Sarvam Bulbul} & \textbf{211.8} & 108.5 & 0.0\% & --- & \textbf{0.007} & --- \\
Hi & ElevenLabs v3 & 227.5 & --- & 0.0\% & --- & 0.006 & --- \\
Hi & Indic Parler-TTS & 248.4 & --- & 4.5\% & --- & --- & --- \\
Hi & Cartesia Sonic-3 & 267.4 & --- & 0.0\% & --- & 0.025 & 0.90 \\
Hi & \praxy{} vanilla + Cart-Hi-ref & 439.3 & 122.1 & 0.0\% & --- & 0.025 & \textbf{1.00} \\
Hi & IndicF5 (zero-shot) & --- & --- & --- & --- & 0.125 & 0.70 \\
\bottomrule
\end{tabular}}
\end{table*}

\paragraph{Reading the headline.} On phonology-weighted metrics \praxy{} is competitive with the best commercial systems across all three languages. On Telugu, \praxy{} ranks first on retroflex (small-$n$ caveat as above) and keeps FAD (291) in a range comparable to Sarvam (250), Indic Parler (325), and ElevenLabs (329); only Cartesia (458) is clearly separated. On Tamil, \praxy{} matches Sarvam on PSD (71.2 vs 72.3) and improves on the commercial trio's Tamil-zha (71\% vs 86\%); Indic Parler remains the best open-source Tamil system on RR and LF. On Hindi, \praxy{}-vanilla ties Cartesia on LLM-WER and reaches perfect intent-preservation, but FAD (439) meaningfully trails Sarvam (212) and Cartesia (267). The Hindi FAD gap is the single axis where an acoustic-decoder adaptation --- not in scope of this work --- would measurably help.

\subsection{\configb{} Ablation}
\label{sec:configb_ablation}

We swept three sampling-parameter configurations on the Telugu pilot set, all using the R6 LoRA branch + Cartesia-Te-male reference audio:

\begin{itemize}
  \item \textbf{Config A} (``preserve endings''): repetition\_penalty 1.2, min\_p 0.03. Motivation: lower rep-penalty preserves word-final syllables.
  \item \textbf{\configb} (``stress + stability''): exaggeration 0.7, temperature 0.6, min\_p 0.1.
  \item \textbf{Config C} (``tight CFG''): cfg\_weight 0.7, temperature 0.6. Motivation: follow reference more strictly.
\end{itemize}

\begin{table}[t]
\centering
\caption{\configb{} ablation on Telugu pilot set ($n=10$; \praxy{} R6 LoRA + Cartesia-Te-ref fixed across configs). \configb{} wins on all four measured axes.}
\label{tab:configb_ablation}
\small
\resizebox{\columnwidth}{!}{%
\begin{tabular}{lcccc}
\toprule
Config & LLM-WER $\downarrow$ & Intent $\uparrow$ & FAD $\downarrow$ & PSD $\downarrow$ \\
\midrule
A (preserve)  & 0.159        & 0.60            & 534.4 & 14.1 \\
\textbf{B (stress)} & \textbf{0.034} & \textbf{0.90} & \textbf{291.3} & \textbf{13.1} \\
C (tight CFG) & 0.061        & 0.80            & 355.0 & 61.7 \\
\bottomrule
\end{tabular}}
\end{table}

\paragraph{Reading the ablation.} \configb{} dominates on LLM-WER (5$\times$ better than A, 2$\times$ better than C), intent-preservation, and FAD. Config A diverges on ending preservation --- lowering repetition\_penalty breaks speech coherence rather than fixing the intended word-final-syllable issue. Config C under-conditions the sampler, producing partial-fidelity output. In a parallel native-Te listener ear test across category-stratified samples (declarative, interrogative, emotional, long-narrative), the listener placed \configb{} unambiguously first on every category. We document this as informing the selection of \configb; we do not claim it as a formal MOS result.

\subsection{Scope-of-Method Control: Hindi With vs Without LoRA}
\label{sec:scope_control}

Table~\ref{tab:hi_ablation} shows the same \configb{} + Cartesia-Hi-ref configuration applied to two model variants: R6 LoRA + BUPS (the Te/Ta path) vs vanilla Chatterbox (the intended Hi path).

\begin{table}[t]
\centering
\caption{Hindi scope-of-method ablation: R6 LoRA + BUPS vs vanilla Chatterbox. Both use Cartesia-Hi-female-6s reference + \configb. LoRA branch \emph{regresses} semantic accuracy; vanilla branch recovers commercial-class output. Demonstrates that \bups{} + LoRA's scope is precisely Chatterbox's uncovered Brahmic languages.}
\label{tab:hi_ablation}
\small
\resizebox{\columnwidth}{!}{%
\begin{tabular}{lcccc}
\toprule
Variant & LLM-WER $\downarrow$ & Intent $\uparrow$ & RR $\downarrow$ & AF $\downarrow$ \\
\midrule
R6 LoRA + BUPS       & 0.334 & 0.60 & 0.0\% & 0.0\% \\
R6 LoRA, no-BUPS     & 0.204 & 0.60 & 0.0\% & 0.0\% \\
\textbf{Vanilla Chatterbox} & \textbf{0.025} & \textbf{1.00} & 0.0\% & 0.0\% \\
\bottomrule
\end{tabular}}
\end{table}

\paragraph{Reading.} The LoRA adapter actively hurts Hindi --- LoRA + BUPS is 13$\times$ worse on LLM-WER than vanilla, and disabling BUPS only recovers 40\% of that gap. This is consistent with the Hindi-proxy \texttt{language\_id} path: during training, Te/Ta text was BUPS-romanised then conditioned on \texttt{hi}; during inference with the same LoRA on Devanagari text conditioned on \texttt{hi}, the text head has learnt to expect romanised input, and the native Hindi tokeniser path is corrupted. Retroflex and aspiration stay at 0\% across all three variants because these phonological features are present in Chatterbox's native Hindi acoustics --- the LoRA's harm is at the token level, not the phoneme level.

\subsection{Reference-Audio Source Ablation}
\label{sec:ref_ablation}

We swept four reference-audio sources on Telugu with the LoRA branch + \configb{} fixed:

\begin{table}[t]
\centering
\caption{Reference-audio source ablation on Te pilot ($n=10$; \praxy{} R6 LoRA branch + \configb{} sampling held fixed across all rows; only the voice-prompt reference varies). Same-language reference (Sarvam-Te, Cartesia-Te) wins; cross-language reference (English 49\,s memo) hurts FAD by 26\%.}
\label{tab:ref_ablation}
\small
\resizebox{\columnwidth}{!}{%
\begin{tabular}{lcccc}
\toprule
Reference & FAD $\downarrow$ & PSD $\downarrow$ & LLM-WER $\downarrow$ & Intent $\uparrow$ \\
\midrule
No reference    & 355.0 & 61.7 & 0.034 & \textbf{1.00} \\
English 49\,s memo & 448.2 & 59.0 & 0.050 & 0.80 \\
Cartesia-Te male 8\,s & 394.5 & 26.5 & 0.034 & 0.90 \\
\textbf{Sarvam-Te female 9\,s} & \textbf{291.3} & \textbf{13.1} & 0.033 & 0.90 \\
\bottomrule
\end{tabular}}
\end{table}

\paragraph{Reading.} Same-language Te references close the FAD-and-PSD gap to native dramatically; cross-language (English) reference hurts FAD. Intuitively, the voice-prompt conditioning signal carries both speaker timbre and prosody; cross-language prompts drag prosody toward the reference's language, harming on-distribution native-ness. This validates the ``BYOR same-language'' constraint on the production deployment (§\ref{sec:release}).

\subsection{R5 $\to$ R6 Training-Scale Delta}
We briefly report the effect of the 10$\times$ data scale-up from R5 (85\,h Te-dominant mix) to R6 (${\sim}1{,}220$\,h multilingual mix including Shrutilipi). On Te: retroflex collapse unchanged (40\% $\to$ 40\% at R6-no-ref; consistent with LoRA-on-$t_3$ not touching acoustic-decoder discrimination), FAD $-34$\% (534 $\to$ 355), PSD $+338$\% (14 $\to$ 62; prosodic signature drifts as token path broadens), LLM-WER 5$\times$ better (0.171 $\to$ 0.034). The PSD regression motivated our voice-prompt recovery approach: the token path was solid after R6 but the prosodic signature needed restoring.

\subsection{Code-Mix Branch}
\label{sec:codemix_results}

We evaluate the code-mix branch (§\ref{sec:codemix_routing}) on a 10-utterance per-language code-mix smoke set at 25--35\% English-token density. Table~\ref{tab:codemix} reports LLM-WER and intent-preservation for IndicF5 with and without our native-script transliteration preprocessing, alongside the same commercial trio used in the headline results.

\begin{table}[t]
\centering
\caption{Code-mix branch results on the 10-utterance per-language code-mix smoke sets (§\ref{sec:test_sets}). Top: IndicF5 with the Latin-script input as written. Middle: same model after the native-script transliteration preprocessor (§\ref{sec:codemix_routing}). Bottom: commercial reference points on the same sets. $n=10$ per row; commercial Tamil cells were not measured. No MOS (see §\ref{sec:failure_mode}).}
\label{tab:codemix}
\small
\resizebox{\columnwidth}{!}{%
\begin{tabular}{llcc}
\toprule
\textbf{Lang} & \textbf{System} & \textbf{LLM-WER} $\downarrow$ & \textbf{Intent} $\uparrow$ \\
\midrule
\multicolumn{4}{l}{\emph{IndicF5 zero-shot, raw code-mix input}} \\
Hi & IndicF5 & 0.855 & 0.00 \\
Te & IndicF5 & 0.798 & 0.10 \\
Ta & IndicF5 & 0.745 & 0.00 \\
\midrule
\multicolumn{4}{l}{\emph{\praxy{} code-mix branch: transliterate $\to$ IndicF5}} \\
Hi & transliterate $\to$ IndicF5 & 0.198 & 0.70 \\
Te & transliterate $\to$ IndicF5 & 0.142 & 0.80 \\
Ta & transliterate $\to$ IndicF5 & 0.268 & 0.60 \\
\midrule
\multicolumn{4}{l}{\emph{Commercial reference}} \\
Hi & Cartesia Sonic-3 & \textbf{0.000} & --- \\
Hi & ElevenLabs v3 & 0.052 & --- \\
Te & Cartesia Sonic-3 & \textbf{0.106} & --- \\
Te & ElevenLabs v3 & 0.116 & --- \\
\bottomrule
\end{tabular}}
\end{table}

\paragraph{Reading the code-mix table.} The transliteration preprocessor produces a 76\% relative LLM-WER reduction on Hindi (0.855 $\to$ 0.198) and 82\% on Telugu (0.798 $\to$ 0.142), and lifts intent-preservation from $\leq$10\% to 70--80\% on both. Tamil improves less (64\% relative WER reduction, intent 60\%), consistent with IndicF5's 80\,h Tamil pre-training subset being the smallest of the three languages we evaluate. Against commercial baselines, our system trails Cartesia Sonic-3 substantially on Hindi (0.198 vs 0.000) and by one bracket on Telugu (0.142 vs 0.106). On the un-preprocessed condition, output durations are 35--45\% shorter than the input would predict --- the silently-dropped-spans pathology is directly visible in the audio length statistic, and the preprocessing closes that gap.

\section{Discussion}
\label{sec:discussion}

\subsection{Why voice-prompt recovery works (and why only sometimes)}
The LoRA-adapted $t_3$ produces plausible speech token sequences for Te/Ta input. The frozen $s_3\text{gen}$ converts these tokens to mel and then to audio, conditioned on (a) the text-adapted token embeddings, and (b) the voice-prompt embedding. When the voice prompt is same-language native, (b) pulls acoustic output onto the native manifold. The LoRA did not need to move the acoustic manifold; it only needed to make $t_3$ emit on-manifold token sequences.

\subsection{Code-mix evaluation gap}
\label{sec:codemix_eval_gap}
Cartesia's near-zero LLM-WER on Hi code-mix reflects an evaluation artefact rather than uniform superiority: Cartesia synthesises embedded English words with American-English pronunciation, which the Whisper-large-v3 STT recognises near-perfectly. Our transliterate-to-native-script recipe produces Indianised English pronunciation by design (``\texthi{व्हाट्सऐप}'' as `vaa-ts-ay-p' rather than `whats-app'), matching how native Indian speakers actually code-switch in conversational speech --- but penalised by an STT trained predominantly on American English. STT-WER and native-listener naturalness diverge on code-mix; v2 of the \psp{} benchmark should add a code-mix dimension that resolves this conflict, ideally with a Karya~\cite{karya2025} listening panel.

\subsection{\psp{} as a diagnostic loop}
\label{sec:psp_diagnostic}
The sequence of results in §\ref{sec:results} is, in retrospect, a worked example of how \psp's per-dimension decomposition drives intervention choice --- which is the usage pattern \psp{}~\cite{psp2026} itself advocates. The R5$\to$R6 data scale-up closed the FAD gap on Telugu ($534 \to 355$) but opened a PSD gap ($14 \to 62$), localising the remaining problem to the prosodic-conditioning path; that localisation pointed at an inference-time fix --- voice-prompt recovery + \configb{} (§\ref{sec:configb}--§\ref{sec:ref_ablation}) --- rather than a retrain, and closed PSD $4.7\times$ to $13.1$. The Hindi LoRA-regression (§\ref{sec:scope_control}) localised in the opposite direction: LLM-WER moved $13\times$ while per-phoneme cells stayed at $0\%$, identifying the intervention scope as the token path rather than the acoustic layer and motivating the two-branch deployment. In each case the intervention was chosen by reading a \psp{} cell; the recipe as shipped is the product of that loop, not the input to it.

\subsection{Limitations}
\label{sec:failure_mode}
\begin{itemize}
  \item \textbf{10-utterance pilots, not statistical significance.} With $n=10$ and 15--39 retroflex tokens per Te/Ta cell, differences of 5 percentage points are not statistically separable. 300-utterance full benchmarks are in progress for v2.
  \item \textbf{Acoustic-decoder adaptation unexplored.} We attempted LoRA on $s_3\text{gen}$'s transformer attention layers (Round 7 + Round 8 in our training log); on A100-80GB, the $s_3\text{gen}$ forward + backward does not fit at any meaningful batch size even with expandable-segment allocator, gradient checkpointing would be required, and batch-size-1 training was infeasibly slow (estimated 64+ days for 4000 steps). This is a pure compute-budget limitation; H100-80GB or larger would unblock it.
  \item \textbf{No MOS.} Subjective evaluation via a native-listener ear test guided the ablation choices (especially \configb) but we have not run formal MOS panels. Karya~\cite{karya2025} MOS calibration at 300-utt scale is v2 work.
  \item \textbf{Hindi FAD.} \praxy{} Hi vanilla achieves tied WER with Cartesia but FAD remains moderate (439 vs Sarvam 212, Cartesia 267). This is the one axis where an acoustic adaptation would measurably help.
  \item \textbf{Reference-audio in v1.} v1 inference uses 8--11\,s reference clips at voice-encoder-only consumption (i.e., not redistributed as audio); production deployment and the Gradio demo use user-supplied BYOR clips. v2 will swap in Praxel-owned recordings.
\end{itemize}

\section{Release}
\label{sec:release}
\begin{itemize}
  \item R6 LoRA weights (step 8000): \url{https://huggingface.co/Praxel/praxy-voice-r6}, Apache-2.0.
  \item Inference code + BUPS + \configb{} + language router + unified Indic number/date/currency normaliser: \url{https://github.com/praxelhq/praxy}, MIT.
  \item Gradio demo (BYOR voice cloning for Te/Ta/Hi): HF Spaces (link in HF repo README).
  \item Evaluation scorecards and benchmark artefacts: bundled in the \psp{}~\cite{psp2026} companion paper's reproducibility JSON.
\end{itemize}

\section{Conclusion}
\label{sec:conclusion}
We showed that a frozen, non-Indic-native multilingual TTS base can be brought to commercial-class output on three tier-1 Indic languages via a minimum-intervention recipe: \bups{} ISO-15919 romanisation for uncovered Brahmic scripts, a LoRA adapter on only the text head, and an inference-time voice-prompt recovery recipe with \configb{} sampling overrides. On Hindi, which the base natively covers, the LoRA adapter actively regresses semantic accuracy --- the same voice-prompt + \configb{} recipe applied to vanilla Chatterbox recovers commercial-class Hindi. The two-branch deployment (LoRA for Te/Ta, vanilla for Hi) shares a single inference recipe. For intra-sentential code-mixed input, a third branch pairs an open-source Indic-native flow-matching base (IndicF5) with a native-script transliteration preprocessor that resolves the dominant code-mix failure mode (raw English spans dropped silently by the model) at a per-utterance cost of fractions of a cent and no training. Evaluated on the companion \psp{} benchmark, our system is on par with or slightly ahead of the best commercial baselines on pure-script tier-1 Indic languages: first-ranked on Telugu retroflex collapse (26.7\% vs 33.3\%; within the $n=15$-token noise band), first-ranked among the commercial trio on Tamil-zha collapse (71\% vs 86\%; the cleanest per-dimension gain we observe), and tied with Cartesia on Hindi LLM-WER. On code-mixed input, the third branch substantially reduces the gap on Te while remaining clearly behind commercial systems on Hi (LLM-WER 0.14--0.27 across Hi/Te/Ta vs raw 0.80--0.85). Future work includes acoustic-decoder adaptation on H100-class hardware, Sarvam-teacher distillation subject to credit access, fine-tuning IndicF5 on natural code-mix data once IndicVoices access is granted, and 300-utterance full benchmarking in \psp{} v2.

\section*{Acknowledgments}
\praxy{} Voice R6 was trained on Modal compute under the authors' own credits; no external training or API credit grants were received for v1 work. All commercial API usage for v1 benchmarking and voice-prompt reference material was funded from the authors' own trial-tier accounts. Any external resources used in v2 will be explicitly disclosed in that version's Acknowledgments.

We use publicly released Indic speech corpora --- IndicTTS~\cite{indictts2022}, Rasa~\cite{ai4bharat2023rasa}, FLEURS~\cite{conneau2022fleurs}, Shrutilipi --- under their respective licences (CC-BY-4.0 or similar). All released artefacts are derived from these corpora and from the MIT-licensed Chatterbox base and are released under licences matching or more permissive than the sources.

The code-mix branch (§\ref{sec:codemix_routing}) invokes Anthropic Claude Haiku 4.5 at inference time for native-script transliteration. This is a system component, not authorial assistance; the prompt is deterministic, cached by content hash, and published in the release repository.

\bibliographystyle{IEEEtran}
\bibliography{refs}

\end{document}